\documentclass[aps,prl,superscriptaddress,floatfix]{revtex4}
\usepackage{amsfonts}
\usepackage{epsfig}
\usepackage{amsmath,amssymb}
\usepackage{times}
\usepackage{xcolor}
\usepackage{multirow}

\usepackage{graphicx}
\usepackage{color}
\usepackage[labelfont=bf,labelsep=period,justification=raggedright]{caption}
\date{}

\begin{document}
\title{Improving the accuracy of the $k$-shell method by removing redundant links-from a perspective of spreading dynamics}

\author{Ying Liu}
\affiliation{Web Sciences Center, University of Electronic
Science and Technology of China, Chengdu 610054, China}
\affiliation{School of Computer Science, Southwest Petroleum University,
Chengdu 610500, China}
\affiliation{Section for Science of Complex Systems, Medical University of Vienna,
Vienna 1090, Austria}

\author{Ming Tang\footnote{Correspondence to: tangminghuang521@hotmail.com}}
\affiliation{Web Sciences Center, University of Electronic
Science and Technology of China, Chengdu 610054, China}

\author{Tao Zhou}
\affiliation{Web Sciences Center, University of Electronic
Science and Technology of China, Chengdu 610054, China}
\affiliation{Big Data Research Center, University of Electronic
Science and Technology of China, Chengdu 610054, China}

\author{Younghae Do}
\affiliation{Department of Mathematics, Kyungpook National
University, Daegu 702-701, South Korea}

\begin{abstract}
Recent study shows that the accuracy of the $k$-shell method in determining node coreness in a spreading process is largely impacted due to the existence of core-like group, which has a large $k$-shell index but a low spreading efficiency. Based on analysis of the structure of core-like groups in real-world networks, we discover that nodes in the core-like group are mutually densely connected with very few out-leaving links from the group. By defining a measure of diffusion importance for each edge based on the number of out-leaving links of its both ends, we are able to identify redundant links in the spreading process, which have a relatively low diffusion importance but lead to form the locally densely connected core-like group. After filtering out the redundant links and applying the $k$-shell method to the residual network, we obtain a renewed coreness $k_S$ for each node which is a more accurate index to indicate its location importance and spreading influence in the original network. Moreover, we find that the performance of the ranking algorithms based on the renewed coreness are also greatly enhanced. Our findings help to more accurately decompose the network core structure and identify influential nodes in spreading processes.

\end{abstract}

\date{\today}
\maketitle


The development of network science has made it a powerful tool to model and analyze complex systems in nature and society~\cite{newman2010}. One fundamental aspect is to understand the complex structures and behaviors of real-world networks~\cite{albert2002,boccaletti2006, dorogov2008,pastor2014}. Network structure can be described from the local, global and meso-scale levels~\cite{fortu2010} such as node degree, clustering, degree distributions, degree correlations, motifs, communities, hierarchies, etc. The $k$-shell decomposition is a method used to partition a network into hierarchically ordered sub-structures ~\cite{Seidman1983}. It decomposes a network in an iterative way, removing all nodes of degree less than current shell index until no removing is possible (see Methods for details). Each node is assigned an index $k_S$ to represent its coreness in the network. Nodes with the same $k_S$ constitute the $k_S$-shell. A large $k_S$ indicates a core position in the network, while a small $k_S$ defines the periphery of the network. The $k$-core, nodes with $k_S\geq k$, obtained in the decomposition process is a highly interconnected substructure in network topology~\cite{dorogovtsev2006}, which has found its application in different fields of science, like biology~\cite{bader2002,wuchty2005,chat2007,peter2009,schwab2010}, economics~\cite{garas2010}, and social science~\cite{wuellner2010,gon2011,reppas2012,colomer2013,garas2014}. For example, nodes in the inner core (large $k_S$ region) have a relatively high probability of being essential and evolutionary conserved in the protein interaction network~\cite{wuchty2005}. Nodes in the innermost core (the shell with the largest $k_S$ value in the network) of the global economic network are most probable to trigger out an economic crisis~\cite{garas2010}. High $k$-cores of the air transportation networks in USA are extremely resilient to both the node removal and edge removal~\cite{wuellner2010}. Because of its low computational complexity of $O(N+E)$~\cite{batagelj2003}, where $N$ is the network size and $E$ is the number of edges in the network, the $k$-shell method is extensively used in analyzing the hierarchical structure of large-scale networks, such as visualizing networks~\cite{hamelin2005}, depicting the network core-periphery features~\cite{holme2005,rossa2013}, and analyzing the Internet and its core~\cite{shai2007,hamelin2008,zhang2008}. In addition, the $k$-core is used to construct network model~\cite{hebert2013}, applied in community detection~\cite{peng2014} and $k$-core percolation is extensively studied which gives a notion of network resilience under random attack~\cite{gleeson2009, baxter2011, cellai2013, zhao2013}. The $k$-shell method is also extended to weighted networks~\cite{marius2013}, dynamic networks~\cite{miorandi2010} and multiplex networks~\cite{Azimi2014}.

Considering that the $k$-shell method decomposes the network into ordered shells from the core to the periphery, researchers found that core nodes of the network are more influential that periphery nodes in a spreading dynamics~\cite{kitsak2010}. Following the work, there is growing interest in using the $k_S$ index to rank nodes of their spreading efficiency. Nodes with large $k_S$ are considered to be more influential and effective than others in a spreading process~\cite{garas2010, gonzalez2011, garas2012, Senpei2014}. Furthermore, some works devise ranking algorithms based on $k_S$ of nodes~\cite{liu2013,bae2014,ren2014,Fu2014}. Despite its effectiveness, however, the coreness determined by the $k$-shell method has some limitations in identifying influential spreaders. In the rumor spreading model, nodes with high coreness are not influential spreaders but act as firewall to prevent the rumor from spreading to the whole network~\cite{borge2012}. For dynamics with steady state, nodes with the highest degree acts more important than the core nodes in uncorrelated networks if the degree distribution of the network has a decay exponent larger than $5/2$~\cite{castellano2012}. In network with tree structure or BA model network, most of the nodes are assigned a same $k_S$ value, thus the $k$-shell index is unable to distinguish node importance~\cite{Senpei2013}. In particular, in our recent study~\cite{liu2015} we shows that in some real-world networks the core nodes as identified by the $k$-shell decomposition are not the most influential spreaders. Specifically speaking, there exists core-like groups which are identified as cores with large $k_S$ but are in fact only locally densely connected groups with relatively low spreading efficiency. This implies that the $k_S$ index may be inaccurate to reflect the location importance of nodes in networks with such local structure, which proposes a great challenge for works using the $k$-shell method to identify network cores and rank nodes.

In this paper, we explore the topological feature of the core-like groups and find out the connection pattern that causes the failure of the $k_S$ index to accurately determine the location importance and spreading influence of nodes in networks with such local structure. Furthermore, we propose a way to improve the accuracy of the $k$-shell method in determining node coreness from the perspective of spreading dynamics. Motivated by the research advances in core-periphery structure~\cite{borgatti1999,peixoto2012,rombach2014,avin2014}, in which core nodes are not only densely connected among themselves but also well connected to the periphery nodes, which are sparsely connected to any other, we consider the characteristics of links a core node should have. Specifically speaking, links of core nodes should not only connect core nodes, but also connect to nodes that are not in the core. To quantitatively determine the effect of a link in a spreading process, we define a measure of diffusion importance based on the connection patterns of its two ends. We find that there exists some redundant links in real-world networks, which have a low diffusion importance but lead to form the core-like group. By filtering out the redundant links from the original network and applying the $k$-shell decomposition on the residual network, we obtain a renewed coreness $k_S$ for each node. This $k_S$ is a much more accurate index to indicate the node importance in a dynamic spreading in the original network. We validate this by simulating the susceptible-infected-recovered (SIR) epidemic process on networks and compare the spreading efficiency of nodes from the core to the periphery, which is used in many research works~\cite{kitsak2010,garas2012}. Furthermore, we find that ranking algorithms based on the $k$-shell method are also greatly enhanced once using the renewed $k_S$ obtained from the residual network.


\section*{Results}
We first present the structural feature of the locally densely connected groups that cause the inaccuracy of the $k$-shell method in determining coreness of nodes in a dynamic spreading. We then define the diffusion importance of edges and remove the redundant edges. Finally, we validate the improved accuracy of the renewed coreness from the perspective of spreading dynamics.

\textbf{Structural feature of locally densely connected group.}
We first focus on six real-world networks in which the $k$-shell method fails to identify the core shells because the existence of the core-like groups~\cite{liu2015}(For the identification of core-like groups, see Methods for details). The properties of the studied networks are listed in Table~\ref{tab:basiccharacteristic}.

\begin{figure}[!ht]
\begin{center}
\epsfig{file=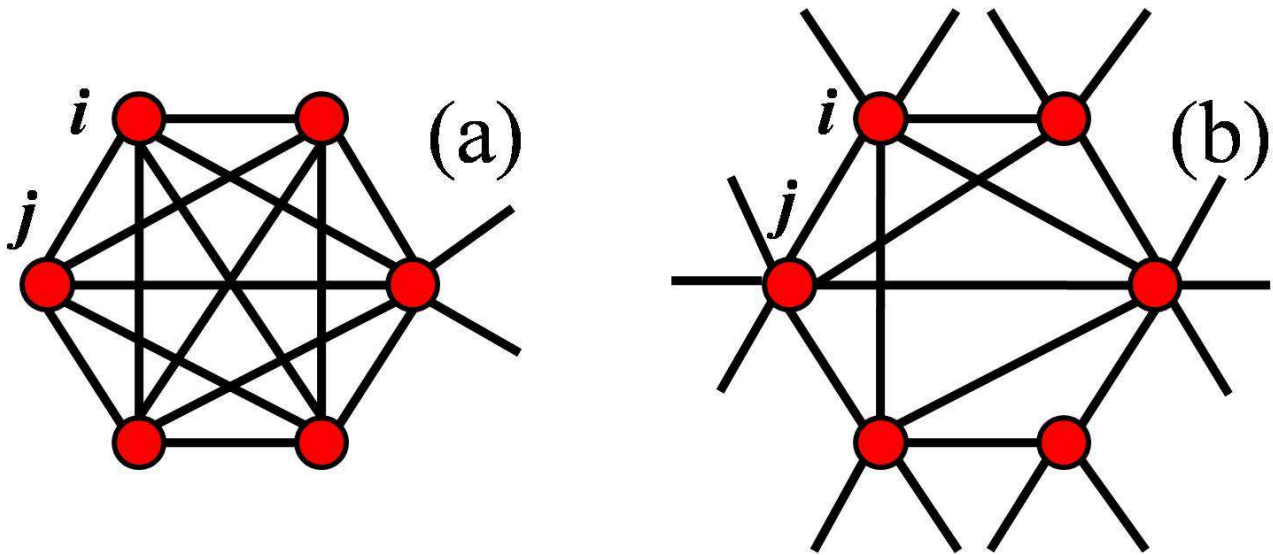,width=1\linewidth}

\caption{\textbf{Illustration of structural feature of the core-like group and the true core.} (a) Core-like group. (b) True core. For the core-like group, core nodes are mutually connected and have very few out-leaving links. While for the true core, core nodes are connected and each of them have a lot of out-leaving links.}
\label{figure1}
\end{center}
\end{figure}

\begin{table*}[!ht]
\caption{\textbf{Properties of the real-world networks studied in this work.} Structural properties include number of nodes ($N$), number of edges ($E$), average degree ($\langle k \rangle$), maximum degree ($k_{max}$), degree heterogeneity ($H_{k}=\langle k^{2} \rangle/\langle k \rangle^{2}$), degree assortativity ($r$), clustering coefficient ($C$), maximum $k_S$ index ($k_{Smax}$), epidemic threshold ($\lambda_c$), infection probability used in the SIR spreading in the main text ($\lambda$) (see Method for details). For the first six networks, there exists core-like groups, while for the last three networks, there is no core-like group in the network, which we will discuss in the last part.}\centering

\begin{tabular}{ccccccccccccc}
\hline
\hline
\textbf{Network} & \textbf{$N$} & \textbf{$E$} & \textbf{$\langle k \rangle$} & \textbf{$k_{max}$} & \textbf{$H_{k}$} & \textbf{$r$} & \textbf{$C$} & \textbf{$k_{Smax}$} & \textbf{$\lambda_c$} & \textbf{$\lambda$}\\
\hline
Email  &1133 &5451 &9.6 &71 &1.942 &0.078 &0.220 &11 &0.06 &0.08\\
CA-Hep &8638 &24806 &5.7 &65 &2.261 &0.239 &0.482 &31 &0.08 &0.12\\
Hamster  &2000 &16097 &16.1 &273 &2.719 &0.023 &0.540 &24 &0.02 &0.04\\
Blog  &3982 &6803 &3.4 &189 &4.038 &-0.133 &0.284 &7 &0.08 &0.27\\
PGP  &10680 &24340 &4.6 &206 &4.153 &0.240 &0.266 &31 &0.06 &0.19\\
Astro  &14845 &119652 &16.1 &360 &2.820 &0.228 &0.670 &56 &0.02 &0.05\\
\hline
Router  &5022 &6258 &2.5 &106 &5.503 &-0.138 &0.012 &7 &0.08 &0.27\\
Emailcontact  &12625 &20362 &3.2 &576 &34.249 &-0.387 &0.109 &23 &0.01 &0.10\\
AS  &22963 &48436 &4.2 &2390 &61.978 &-0.198 &0.230  &25 &0.004 &0.13\\
\hline
\hline
\end{tabular}
\label{tab:basiccharacteristic}
\end{table*}

Based on in-depth analysis of the network local structure, we find that the core-like group has a clique-like local structure as shown in Fig.~\ref{figure1} (a). Most of the nodes in the core-like group have a similar connection pattern. Let's take node $i$ for example. Neighbors of node $i$ are mutually connected, with only one neighbor having a few out-leaving link, that are links connecting outside the neighborhood of node $i$. In the $k$-shell decomposing process, node $i$ will be assigned a $k_S$ value equal to its degree. Considering the feature of core in the core-periphery structure~\cite{borgatti1999,rombach2014}, which is densely connected among themselves and well connected to the periphery, we think that the cohesive group shown in Fig. 1 (a) is not a true core, because it is only densely connected within a group but not well connected to the remaining part of the network. When a disease origins from node $i$, most of the infections are limited in the neighborhood of node $i$. As for the true core in Fig.1 (b), core nodes are well connected and at the same time connect well to the outside of the core. When a disease or rumor origins from node $i$, it is easier to spread to a broad area of the network through neighbors of node $i$ whose links are connecting to the external parts of $i$'s neighborhood. We take the innermost core of the network CA-Hep and Router for example and visualize the connection pattern of the innermost core by the software Gephi of version 0.8.2~\cite{bastian2009}. We find that for the innermost core of CA-Hep, which is the $31$-shell composed of 32 mutually connected nodes, has a structure very similar to the structure shown in Fig. 1 (a), with only five nodes having a small number of links out leaving the group, as shown in Fig. S1 (a) in Supporting Information (SI). As for the innermost core of Router, which is the $7$-shell composed of 26 nodes, each nodes connects well to a large amount of nodes that are not in the core-shell, as shown in Fig. S1 (b). Motivated by the structural difference of the core-like group and the true core, we think that the importance of links of a node $i$ varies depending on the connection pattern of its neighbor nodes (e.g. node $j$): if node $j$ has many connections out-leaving node $i$'s neighborhood, the probability of infecting more nodes increases when the spreading origins from node $i$, and thus the edge linking node $i$ and node $j$ is important for node $i$. On the other hand, if node $j$ has very few or even no out-leaving links from node $i$'s neighborhood, the probability of infecting a large population by node $i$ decreases, and thus the edge linking node $i$ and node $j$ is less important.

\begin{figure}[!ht]
\begin{center}
\epsfig{file=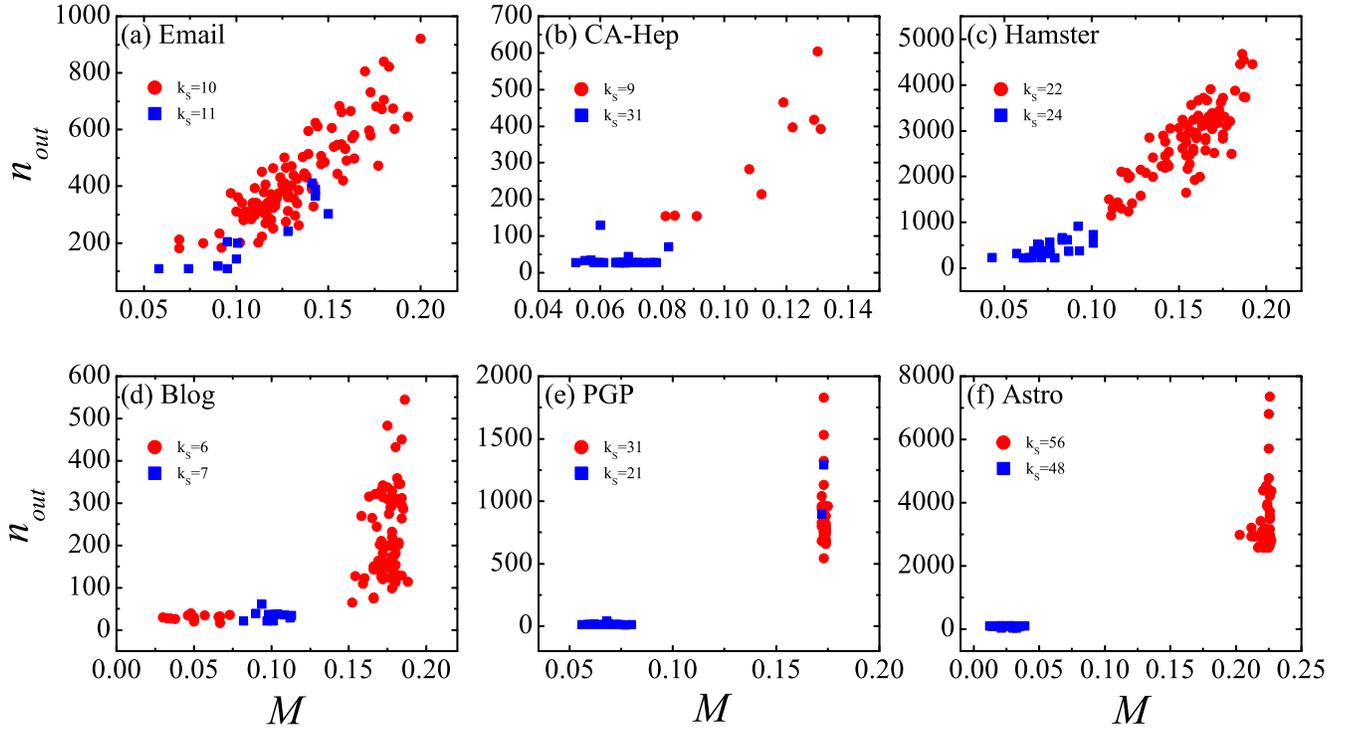,width=1\linewidth}

\caption{\textbf{Correlation of spreading efficiency and the number of out-leaving links.} For each network, we present the nodes in the core-like group (blue squares) and in the highest spreading efficiency shell (red circles). A positive correlation between the spreading efficiency and the number of out-leaving links is demonstrated.}
\label{figure2}
\end{center}
\end{figure}

To confirm the relationship between the structure feature and spreading behavior on it, we use the SIR spreading model~\cite{anderson1991} to simulate the spreading process on networks. We record the spreading efficiency of each node, which is the size of the final infected population $M$ when a spreading origins from the node (see Methods for details). Then we study the correlation between the total number of out-leaving links $n_{out}$ of a node, that is the sum of out-leaving links over all neighbors the node, and its spreading efficiency $M$. To find out the difference between the core-like group and the true core, we choose two groups of nodes for each networks. The first one is the shell that is a core-like group (there may be several core-like groups in the network, and we choose the one with the largest $k_S$ value); the second one is the shell with the highest average spreading efficiency. From Fig.~\ref{figure2} we can see that in general nodes in core-like groups (blue squares), which have a relatively low spreading efficiency, have a lower number of out-leaving links than nodes in the highest spreading shell (red circles). What is worth noticing is that although most nodes in core-like groups have a relatively low spreading efficiency, there may be some nodes that have a high spreading efficiency, corresponding to some blue nodes in Email and PGP, which also have a relatively high number of out-leaving links. On the other hand, in the highest spreading efficiency shell, there are nodes with relatively low spreading efficiency whose number of out-leaving links is correspondingly low, such as some red nodes in Email and Blog. These indicate a positive correlation between the spreading efficiency and the number of out-leaving links of a node through its neighbors.

Considering the structural feature of core-like groups and the correlation between the number of out-leaving links and the spreading efficiency of a node, we realize that in the locally densely connected structures, there exists some links which lead to form a clique-like local structure but contribute little to the spreading process. This causes the failure of $k$-shell method in accurately determining nodes coreness and identifying true cores in many real-world networks, from the perspective of spreading efficiency. Next we will step further to find a way to eliminate the negative effect of these links and improve the accuracy of the $k$-shell method in determining network core structure.

\textbf{Defining the diffusion importance for links.} We define the diffusion importance of links in the following way. Consider an edge $e_{ij}$. When a disease spreads along it, there are two possible directions. In one direction, the disease origins from node $i$ and spreads along $e_{ij}$ to node $j$, and then spreads to the other parts of the network through node $j$. We record the number of links of node $j$ connecting  outside the nearest neighborhood of node $i$ as $n_{i\rightarrow j}$. In the other direction, the disease origins from node $j$ and spreads along $e_{ji}$ (the same edge as $e_{ij}$ because it is undirected edge) to node $i$, and then spreads through node $i$ to the other parts of the network. We record the number of links of node $i$ connecting outside the nearest neighborhood of node $j$ as $n_{j\rightarrow i}$. Then the diffusion importance of edge $e_{ij}$ is defined as
\begin{equation}
D_{ij}=(n_{i\rightarrow j}+n_{j\rightarrow i})/2.
\end{equation}
This value quantifies the average potential influence of an edge in both directions. Let's take edge $e_{ij}$ in Fig. 1 as an example to calculate the diffusion importance. In Fig.1 (a), $n_{i\rightarrow j}=0$, which is the number of links of node $j$ that connect outside the neighborhood of node $i$. At the same time, $n_{j\rightarrow i}=0$, which reflects that node $i$ has no links connecting to nodes that are not in the neighborhood of node $j$. Thus the $D_{ij}=0$. In Fig. 1 (b), $n_{i\rightarrow j}=3$, $n_{j\rightarrow i}=2$, and thus $D_{ij}=2.5$. In this way, we can calculate the diffusion importance for all edges in the network. When each edge is assigned a diffusion importance, the unweighted graph becomes weighted graph. The weight on edge contains the information of the potential spreading coverage when a disease spreads along the edge. For a general discussion of the weighted network is not in the scope of this paper, which we will explore in the future. Here, we concentrate on identifying links that is less important in the spreading process but lead to form a locally densely connected local structure, which results in the failure of the $k$-shell method to accurately determine the coreness of nodes in spreading dynamics.

\textbf{Filtering out redundant links and applying the $k$-shell method to obtain a new coreness for nodes.}
From the analysis of Fig.1, we come to the idea that links with low diffusion importance are redundant links, which contribute much to a densely connected local structure and a high $k_S$ for nodes but have a limited diffusion influence. We set a redundant threshold $D_{thr}$ to determine redundant links. If $D_{ij}< D_{thr}$, edge $e_{ij}$ is considered as a redundant link. If we use $G=\{V, E\}$ to represent a graph, where $V$ is the set of nodes and $E$ is the set of edges, then the residual network that is obtained by filtering out redundant links is represented as $G^{\prime}=\{V^{\prime}, E^{\prime}\}$, where $V^{\prime}=V$ and $E^{\prime}\subseteq E$. If all edges in the network have a $D_{ij}\geq D_{thr}$, then $E^{\prime}=E$.

We first apply the $k$-shell decomposition to the original networks and obtain the coreness for each node, recorded as $k_S^{o}$. Then we identify and filter out the redundant links. Given that filtering out too many edges may destruct the main structure of the network, the $D_{thr}$ should not be too large which will lead to a large proportion of links being identified as redundant links. Meanwhile, the $D_{thr}$ should not be too small because the redundant links that contribute much to the local densely connected structure may have a diffusion importance greater than 0 but are still not so important in a spreading process. We adopt a diffusion threshold of $D_{ij}=2$. For a discussion of the diffusion threshold, please see SI for details. In this case, edges with $D_{ij}\geq 2$ are remained in $G^{\prime}$. We apply the $k$-shell method to $G^{\prime}$ and obtain a renewed coreness for each node, recorded as $k_S^{r}$. We use the imprecision function, which is initially proposed by Kitsak \textit{et al.}~\cite{kitsak2010} and modified by Liu \textit{et al.}~\cite{liu2015}, to compare the accuracy of $k_S^{o}$ and $k_S^{r}$ in determine node coreness in the network. The imprecision function is defined as
\begin{equation}
\varepsilon(k_S)=1-\frac{M_{core}(k_S)}{M_{eff}(k_S)},
 \end{equation}
where $k_S$ is the variable ranging from $0$ (for isolated nodes in the residual network) to the maximum $k_S$ value in the network. $M_{core}(k_S)$ is the average spreading efficiency of nodes with coreness $k_S'\geq k_S$ (nodes in $k_S$-core), and $M_{eff}(k_S)$ is the average spreading efficiency of $n$ nodes with the highest spreading efficiency, where $n$ equals to the number of nodes in $k_S$-core.  This function quantifies how close to the optimal spreading is the average spreading of nodes in $k_S$-core. A small $\varepsilon(k_S)$ value means nodes identified as in core shells have a correspondingly high spreading efficiency.

\begin{figure}[!ht]
\begin{center}
\epsfig{file=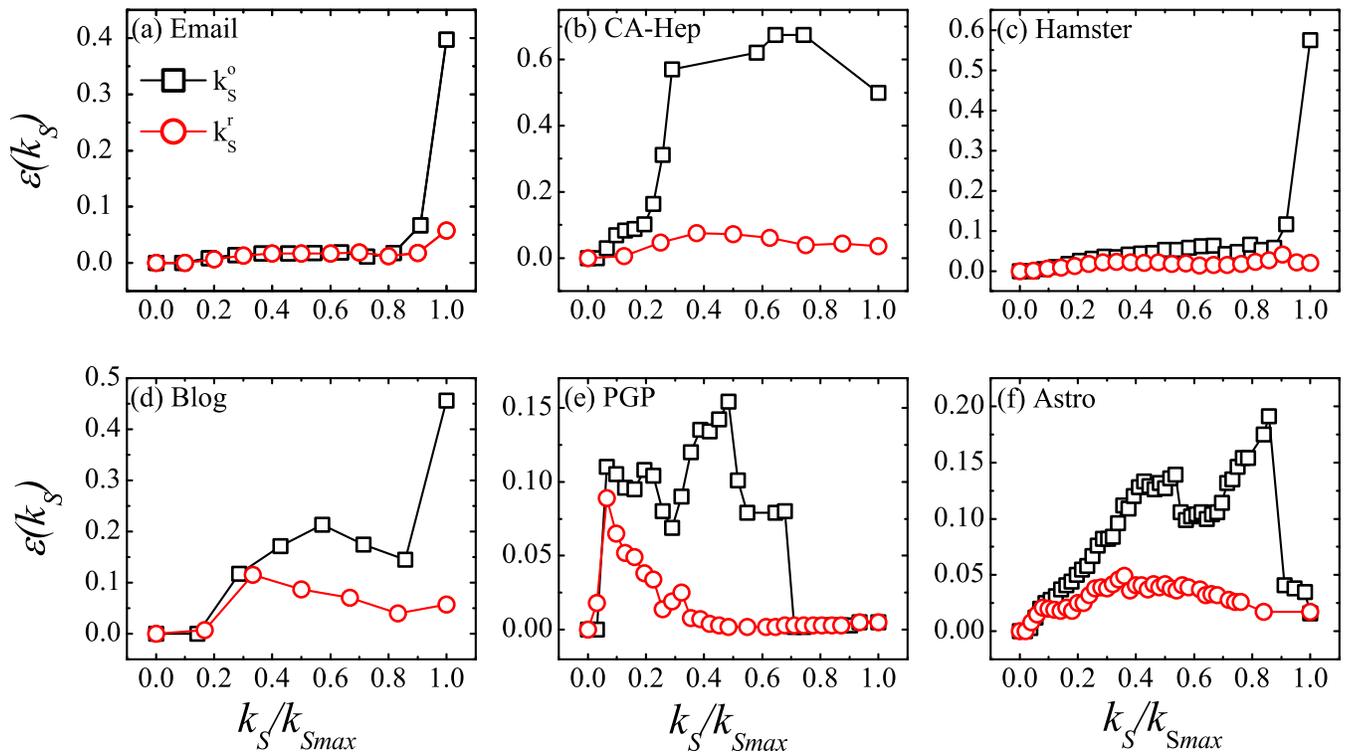,width=1\linewidth}

\caption{\textbf{The imprecision of $k_S^{o}$ and $k_S^{r}$ as a function of shell index.} $k_S^{o}$ is the coreness obtained from the original network, and $k_S^{r}$ is the coreness obtained from the residual network. Shell index $k_S$ ranges from 0 to $k_{Smax}$ and is normalized by $k_{Smax}$. The imprecision of $k_S^{r}$ is obviously smaller than that of $k_S^{o}$.}
\label{figure3}
\end{center}
\end{figure}

In Fig.\ref{figure3} we compare the imprecision of $k_S^{o}$ and $k_S^{r}$. The number of shells may be different for the original graph $G$ and the residual graph $G^{\prime}$, so we normalized the shell index $k_S$ by the maximum shell index $k_{Smax}$ in $G$ and $G^{\prime}$ respectively. The imprecision based on $k_S^{r}$ is in general obviously lower than the imprecision based on $k_S^{o}$. For the networks of Email, CA-Hep, Hamster and Blog, the imprecision of $k_S^{o}$ is high for large values of $k_S$, close to or above $0.4$. This means that in these networks nodes identified as core by the $k_S^{o}$ are in fact not very influential in a spreading process. In the network of PGP and Astro, there are sudden jumps in $k_S^{o}$ imprecision, which correspond to the locally densely connected structure that does not exist in the innermost core but exist in some outer shells of the network~\cite{liu2015}. On the contrary, when the $k_S^{r}$ is used to determine node coreness, a much lower imprecision is obtained. In all the studied real-world networks, the absolute value of the imprecision function based on $k_S^{r}$ is close to or smaller than 0.1. This means that the $k_S^{r}$ is a good indicator of spreading efficiency. After removing the redundant links with low $D_{ij}$ values, the accuracy of the $k$-shell method in determining cores is greatly greatly improved.

In many cases, people are more interested in top ranked nodes, which corresponds to leaders in the society. We rank nodes by their coreness $k_S^{o}$ and $k_S^{r}$ respectively and compare the accuracy of coreness in identifying the most influential spreaders. Results show that the coreness obtained from the residual network is much more accurate than the original coreness in identifying the most influential spreaders. See Fig. S3 in SI for more details.

\begin{figure}[!ht]
\begin{center}
\epsfig{file=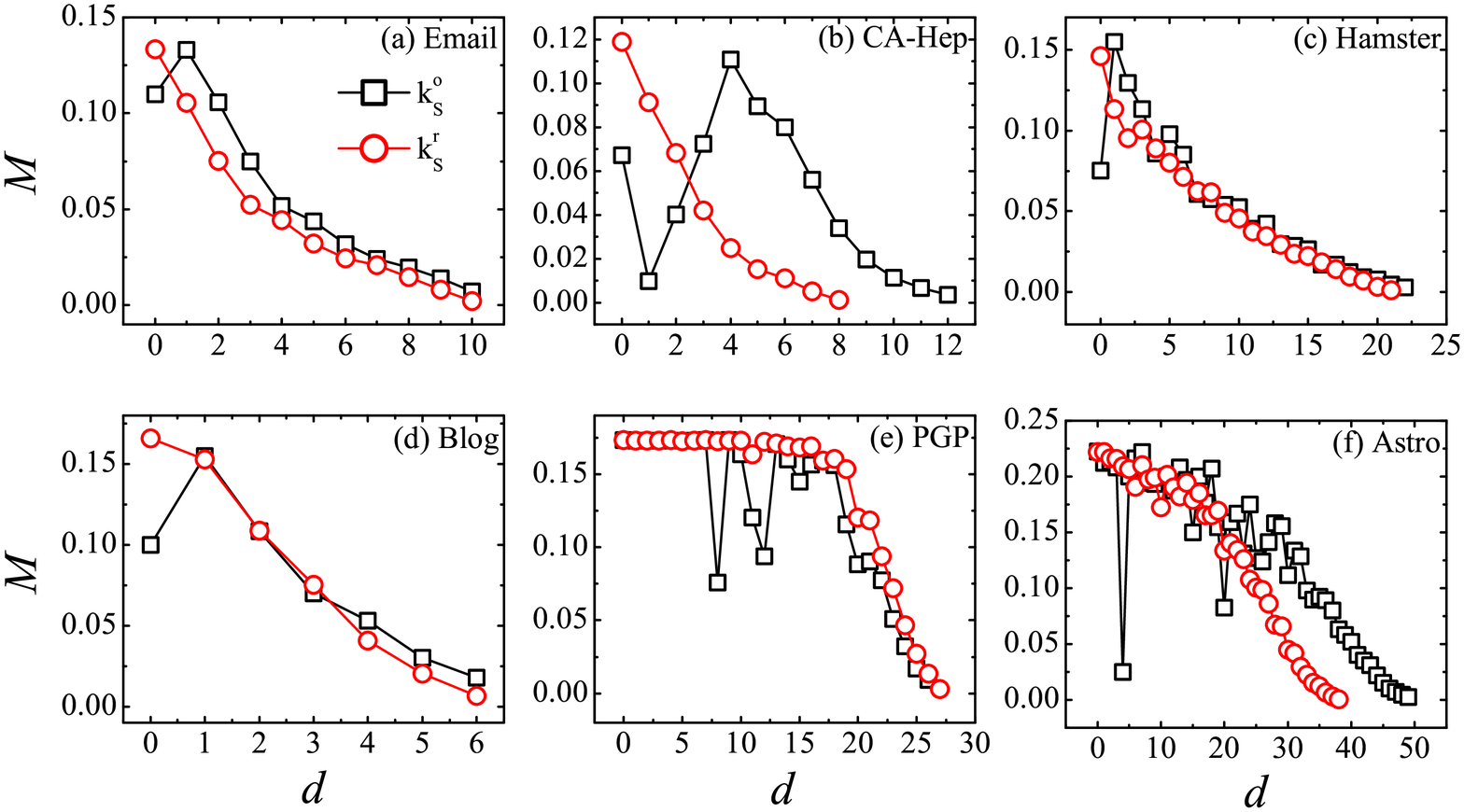,width=1\linewidth}
\caption{\textbf{Spreading efficiency of a shell and its distance from the innermost core.} $k_S^{o}$ is the coreness obtained from the original network, and $k_S^{r}$ is the coreness obtained from the residual network. $d$ is the distance from the innermost core. $d=0$ corresponds to the innermost core. }
\label{figure4}
\end{center}
\end{figure}

Then we focus on the spreading efficiency of shells. A good partition of the network is supposed to display a concordant trend between the shell index obtained from network topology and the spreading efficiency of that shell. One would expect that shells with large $k_S$ should have a higher spreading efficiency than shells with small $k_S$. We plot the spreading efficiency $M$ of each shell (expressed as the distance $d$ of a shell from the innermost core), where the spreading efficiency of a shell is the average spreading efficiency of nodes in that shell. As shown in Fig.~\ref{figure4}, the spreading efficiency of shells are in general decreasing monotonically with the increase of distance from the innermost core in all studied networks when $k_S^{r}$ is used. In the networks of Email, CA-Hep and Blog, the spreading efficiency of each shell and its coreness $k_S^{r}$ is completely concordant. A large $k_S^{r}$ indicates a higher spreading efficiency of the shell. In the networks of Hamster, PGP and Astro, the spreading efficiency and its coreness $k_S^{r}$ are concordant in most shells. There are a limited number of shells where the trend is not so monotonic, however the fluctuation in spreading efficiency is relatively small compared to that of the $k_S^{o}$. As for the $k_S^{o}$, the trend is not as monotonic as $k_S^{r}$. In other words, the coreness obtained from the residual network predicts the spreading efficiency much more accurate than the original one.

\textbf{Comparing with random deletion and other way of targeted removing of links.} Our way of removing redundant links obviously improve the accuracy of the $k$-shell method in determining the influence of nodes in a spreading. Now we compare the effectiveness of our way of targeting the redundant links with random deletion, as well as targeting links whose importance is determined by the degree of nodes on its two ends. To compare with random deletion, we randomly select a set of edges and delete them from the network. The number of edges that is to be deleted is the same as the number of identified redundant links. Then we apply the $k$-shell decomposition to the residual network and obtain a $k_S$ for each node. We realize the random deletion for $50$ times and average the $k_S$ obtained at each realization as coreness for node $i$, which we record as $k_S^{a}$ to represent random or arbitrary deletion. A comparison of the imprecision as a function of shell index are shown in Fig. \ref{figure5}. In most cases, the imprecision of $k_S^{a}$ is very close to that of the $k_S^{o}$ obtained from the original network, and is obviously higher than the imprecision of $k_S^{r}$ obtained from the residual network. This implies that the core-like groups still exist in the residual network after random deletion of links. Although the imprecision of $k_S^{a}$ is slightly improved in some networks, we think it is because that when the links are selected randomly, there is a chance that a redundant link is selected.

A widely used way of determining edge importance is considering the degree of nodes on its two ends. The weight (also the importance) of an edge $e_{ij}$ is proportional to the product of $k_{i}$ and $k_{j}$ as $w_{ij}={(k_{i}k_{j})}^{\theta}$, where $k_{i}$ and $k_{j}$ are the degree of node $i$ and node $j$ respectively~\cite{barrat2004, wang2008, tang2011} and $\theta$ is a tunable parameter. This measure is also strongly correlated with the betweenness centrality of an edge~\cite{holme2002}. We use a parameter $\theta=1$ to determine the edge importance, and remove the edges of small weight from the network to see its effect on the $k$-shell method. The number of edges removed is the same as the number of redundant links identified. We find that the imprecision of coreness $k_S^{w}$ obtained from the residual network in this way is almost the same as the original $k_S^{o}$, as shown in Fig. S4 in SI.

The above analysis implies us two points. First, our way of identifying and removing the redundant links is effective in improving the accuracy of $k$-shell method in profiling the core structure of the network from the perspective of spreading dynamics. Second, the $k$-shell index has a robustness against random failure, which is consistent with the result in Ref. \cite{kitsak2010}. In that work, authors pointed out that the $k$-shell method is robust under random deletion of even up to 50\% of the edges, which means the relative ranking of the $k_S$ value for the same nodes in the original network and the network after random deletion are almost the same.

\begin{figure}[!ht]
\begin{center}
\epsfig{file=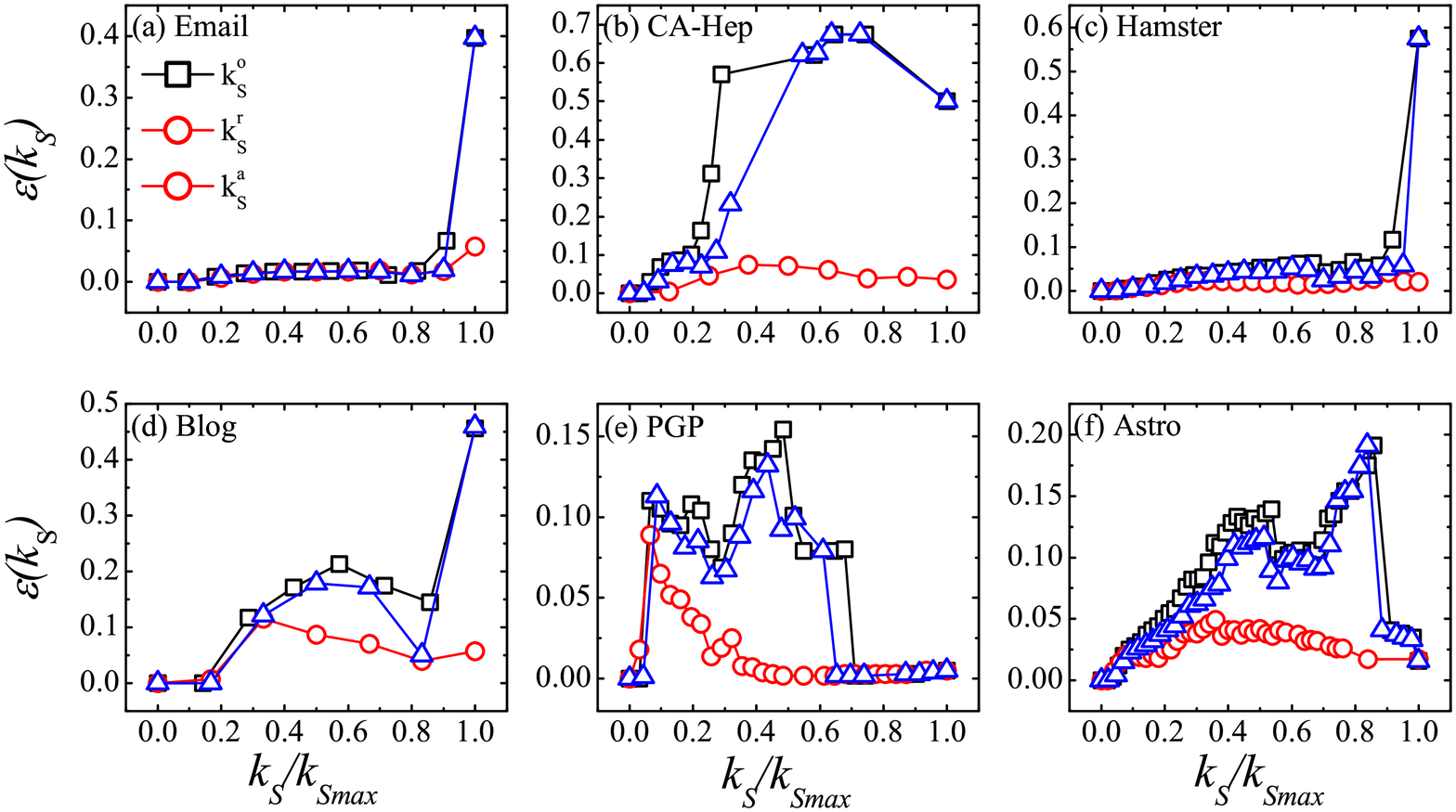,width=1\linewidth}
\caption{\textbf{The imprecision of $k_S^{o}$, $k_S^{r}$ and $k_S^{a}$ as a function of shell index.} $k_S^{o}$ is the coreness obtained from the original network, $k_S^{r}$ is the coreness obtained from the residual network and $k_S^{a}$ is the coreness obtained from the network after random deletion of edges. Shell index $k_S$ ranges from 0 to $k_{Smax}$ and is normalized by $k_{Smax}$. The imprecision of $k_S^{r}$ is obviously smaller than that of $k_S^{o}$ and $k_S^{a}$.}
\label{figure5}
\end{center}
\end{figure}

\section*{Discussion}
Profiling the network hierarchical structure is very important in understanding the behaviors on it. The $k$-shell decomposition is a basic method to describe network structure and identify core areas that is used in many fields of science. We study the $k$-core structure of real-world networks and the spreading process on it. We find that the accuracy of the $k$-shell method in identifying influential spreaders is impacted by the locally densely connected group in the network, which correspond to real-world scenarios such as extensive communication and cooperation within a small group or community. Based on in-depth analysis of network local structure and motivated by research advances in core-periphery structure, we realize that the links of nodes contribute differently to the affected population in a spreading process. For the first time we define a diffusion importance for each link in the network based on its potential influence in a spreading process. By filtering out redundant links and then applying the $k$-shell decomposition to the residual graph, we get a renewed coreness for nodes. Experimental results show that this renewed coreness is much more accurate in determining the spreading influence of node from the core to the periphery. Specifically speaking, the imprecision of coreness in identifying influential spreaders is greatly reduced. Nodes with high renewed coreness are in general have a higher spreading efficiency than nodes with low renewed coreness.

There are many algorithms using the $k_S$ index as a global importance of nodes and ranking nodes. Among them, the iterative resource allocating (IRA) algorithm~\cite{ren2014} greatly enhance the accuracy of centrality measures in ranking node influence by iteratively relocating sources to each node based on the centrality of its neighbors (see Methods for details). After iteration, the resource of a node will be stable and is used to rank node of its spreading influence. As above, we filter out the redundant links of $G$ and apply the $k$-shell decomposition to the residual graph $G^{\prime}$ to obtain a $k_S^{r}$ and then implement the IRA algorithm on $G^{\prime}$. We find that the ranking accuracy is greatly improved, as shown in Fig. S5. The effectiveness of our method in another ranking algorithm, which defines a neighborhood coreness $C_{nc}$ of node $i$ as $C_{nc}=\sum_{j\in \Gamma(i)}k_S(j)$ in~\cite{bae2014}, where $\Gamma(i)$ is the set of neighbors of node $i$ and $k_s(j)$ is the coreness of node $j$, is shown in SI Fig. S6. We still come to a great improvement in the ranking accuracy.

As our way of filtering out redundant links works well for networks with locally densely connected structure, one may ask the performance of $k_S^{r}$ on networks with no such local structure. For the networks of Router, Emailcontact and AS listed in Table 1, in which there is no core-like group and the $k$-shell method works well on the original network, we find that by filtering out redundant links, the performance of $k_S^{o}$ and $k_S^{r}$ are nearly exactly the same, implying that there is no negative effect on the $k$-shell method on networks where it works well. We present the coreness imprecision as a function of shell index and percentage of nodes $p$ in SI Fig. S7 and S8 respectively, as well as the spreading efficiency of each shell in Fig. S9. It is again due to the robustness of the $k$-shell method. This feature is meaningful in that our way of filtering out redundant links will greatly improve the accuracy of the $k$-shell method in networks where it doesn't work well while at the same time do not impact its performance in networks where it already works well. We also test the effects of filtering out redundant links on other centrality measures such as degree centrality, betweenness centrality and eigenvector centrality in ranking node's spreading influence. Results show that the ranking performance of the centrality obtained from the residual network remains very close to the centrality obtained from the original network. This means the redundant links has little influence on these centrality measures, which is a proof of the redundancy of these links.

The identification of redundant links gives us implication that redundancy has an impact on the analysis of network structure. While we only concentrate on its effectiveness in $k$-shell method and from the perspective of spreading dynamics, the influence of redundant links on other network analysis remains unexplored, such as community partition and network controllability. This propose two challenges. First, we need to decide which structural features of network are affected much by redundant links. Second, how to define the importance of links in the network may depend on the behaviors on it such as rumor spreading, synchronization and immunization. In addition, while our way of determining the redundant threshold $D_{thr}$ is obtained from simulation experiments, a parameter-free way of identifying the redundant links is worthy of further explore.

\section*{Methods}
\textbf{The $k$-shell decomposition.}
The algorithm starts by removing all nodes with degree $k=1$. After removing all nodes with $k=1$, there may appear some nodes with only one link left. We iteratively remove these nodes until there is no node left with $k=1$. The removed nodes are assigned with an index $k_S=1$ and are considered in the 1-shell. In a similar way, nodes with degree $k\leqslant2$ are iteratively removed and assigned an index $k_S=2$. This pruning process continues removing higher shells until all nodes are removed. Isolated nodes are assigned an index $k_S=0$. As a result, each node is assigned a $k_S$ index, and the network can be viewed as a hierarchical structure from the innermost shell to the periphery shell.

\textbf{Identify core-like groups in real-world networks.}
The link entropy of a shell with index $k_S$ is defined~\cite{liu2015} as
\begin{equation}\label{entropy}
H_{k_S}=-\frac{1}{lnL}\sum_{k'_S=1}^{k_{Smax}}r_{k_S,k'_S}lnr_{k_S,k'_S},
\end{equation}
where $r_{k_S,k'_S}$ is the average link strength of nodes in the $k_S$-shell to the $k'_S$-shell and $L$ is the number of shells in the network. The link strength of node $i$ to the $k'_S$-shell is the ratio of the number of links originating from node $i$ to the shell with index $k'_S$ to the total number of links of node $i$. The shells which have a relatively low entropy compared with its adjacent shells are usually locally connected core-like groups.

\textbf{SIR model}.
We use the susceptible-infected-recovered (SIR) spreading model to simulate the spreading process on networks and obtain the spreading efficiency for each node. In the model, a node has three possible states: $S$ (susceptible), $I$ (infected) and $R$ (recovered). Susceptible individual become infected with probability $\lambda$ if it is contacted by an infected neighbor. Infected nodes contact their neighbors and then they change to recovered state with probability $\mu$. For generality we set $\mu=1$. Recovered nodes will neither be infected any more nor infect others, and they remain the $R$ state until the spreading stops. Initially, a single node is infected and all others are susceptible. Then the disease spreads from the seed node to the others through links. The spreading process stops when there is no infected node in the network. The proportion of recovered nodes $M$ when spreading stops is considered as the spreading capability, or spreading efficiency, of the origin node. We realize the spreading process for $100$ times and take the average spreading efficiency of a node as its spreading efficiency.

As we have discovered that the infection probability will not change the relative spreading efficiency of nodes, we chose an infection probability $\lambda>\lambda_{c}$, where $\lambda_{c}=\langle k\rangle/(\langle k^{2}\rangle-\langle k\rangle)$ is the epidemic threshold determined from the heterogenous mean-field method~\cite{castellano2010}. Under the infection probability of $\lambda$, the final infected population $M$ is above $0$ and reaches a finite but small fraction of the network size for most nodes as spreading origins, in the range of $1\%$-$20\%$~\cite{kitsak2010}.

\textbf{Ranking algorithm of IRA}.
This algorithm considers that the spreading influence of a node is determined by both its centrality and its neighbor's centrality~\cite{ren2014}. In an iterative resource allocation process, the resource of nodes is distributed to its neighbors according to their centrality. The resource node $i$ receive is
\begin{equation}
I_i(t+1)=\sum_{j\in\Gamma_{i}}R_{j\rightarrow i}(t+1)=
\sum_{j\in\Gamma_{i}}(\frac{\theta_i^\alpha}{\sum_{u\in\Gamma_{j}}\theta_u^\alpha}\delta_{ij})I_j(t),
\end{equation}
where $R_{j\rightarrow i}(t+1)$ is the amount of resource distributed from node $j$ to node $i$ at time t+1, $\Gamma_{i}$ is the sets of node $i$'s neighbors. $\theta_i$ is the centrality of node $i$, and $\alpha$ is a tunable parameter to adjust the influence of centrality. $u$ belongs to the neighborhood $\Gamma_{j}$ of node $j$. $\delta_{ij}=1$ if there is a link between node $i$ and node $j$, otherwise $\delta_{ij}=0$. $I_j(t)$ is the resource hold by node $j$ at time step t. Initially, each node has an unit resource. The resource distributed to each node will be stable after iterations, and the final resources of nodes are used to rank their spreading influence. The coreness centrality is used here, and $\alpha$ is set to 1.

\textbf{Data sets.}
The real networks studied in the paper are:  (1) Email (e-mail network of University at Rovira i Virgili, URV) ~\cite{guimera2003};(2) CA-Hep (Giant connected component of collaboration network of arxiv in high-energy physics theory)~\cite{leskovec2012}; (3) Hamster (friendships and family links between users of the website hamsterster.com)~\cite{hamster2014}; (4) Blog (the communication
relationships between owners of blogs on the MSN (Windows Live) Spaces website)~\cite{xie2006};  (5) PGP (an encrypted communication network)~\cite{boguna2004}; (6) Astro physics (collaboration network of astrophysics scientists)~\cite{newman2001}; (7) Router (the router level topology of the Internet, collected by the Rocketfuel Project)~\cite{spring2004};(8) Email-contact (Email contacts at Computer Science Department of University college London)~\cite{kitsak2010}; (9) AS (Internet at the autonomous system level)~\cite{newmandataas}.

\section*{Figure legends}

{\bf Figure 1}: Illustration of structural feature of the core-like group and the true core. (a) Core-like group. (b) True core. For the core-like group, core nodes are mutually connected and have very few out-leaving links. While for the true core, core nodes are connected and each of them have a lot of out-leaving links.

{\bf Figure 2}:
Correlation of spreading efficiency and the number of out-leaving links. For each network, we present the nodes in the core-like group (blue squares) and in the highest spreading efficiency shell (red circles). A positive correlation between the spreading efficiency and the number of out-leaving links is demonstrated.

{\bf Figure 3}:
The imprecision of $k_S^{o}$ and $k_S^{r}$ as a function of shell index. $k_S^{o}$ is the coreness obtained from the original network, and $k_S^{r}$ is the coreness obtained from the residual network. Shell index $k_S$ ranges from 0 to $k_{Smax}$ and is normalized by $k_{Smax}$. The imprecision of $k_S^{r}$ is obviously smaller than that of $k_S^{o}$.

{\bf Figure 4}: Spreading efficiency of a shell and its distance from the innermost core. $k_S^{o}$ is the coreness obtained from the original network, and $k_S^{r}$ is the coreness obtained from the residual network. $d$ is the distance from the innermost core. $d=0$ corresponds to the innermost core.

{\bf Figure 5}: The imprecision of $k_S^{o}$, $k_S^{r}$ and $k_S^{a}$ as a function of shell index. $k_S^{o}$ is the coreness obtained from the original network, $k_S^{r}$ is the coreness obtained from the residual network and $k_S^{a}$ is the coreness obtained from the network after random deletion of edges. Shell index $k_S$ ranges from 0 to $k_{Smax}$ and is normalized by $k_{Smax}$. The imprecision of $k_S^{r}$ is obviously smaller than that of $k_S^{o}$ and $k_S^{a}$.

\section*{Acknowledgement}
This work was partially supported by the National Natural Science Foundation of
China (Grant Nos. 11105025, 61433014), the Scientific Research Starting Program of Southwest Petroleum University (No. 2014QHZ024), the Chinese Scholarship Council under No. 201406070071, and the National Research Foundation of Korea (NRF-2013R1A1A2010067).

\section*{Author contributions}
Y. L., M. T., T. Z. and Y.H.D devised the research project.
Y. L. implements experiments.
Y. L. and M. T. analyzed the results.
Y. L., M. T., T. Z. and Y.H.D wrote the paper.

\section*{Additional information}



{\bf Competing financial interests}:
The authors declare no competing financial interests.


~\\

\begin{center}

{\Large
Supporting Information for\\
\vspace{0.5cm}
\textbf{Improving the accuracy of the $k$-shell method by removing redundant links-from a perspective of spreading dynamics}
}\\
\vspace{0.5cm}

\large{Ying Liu, Ming Tang, Tao Zhou and Younghae Do}

\end{center}

\begin{figure}[!ht]
\begin{center}
\epsfig{file=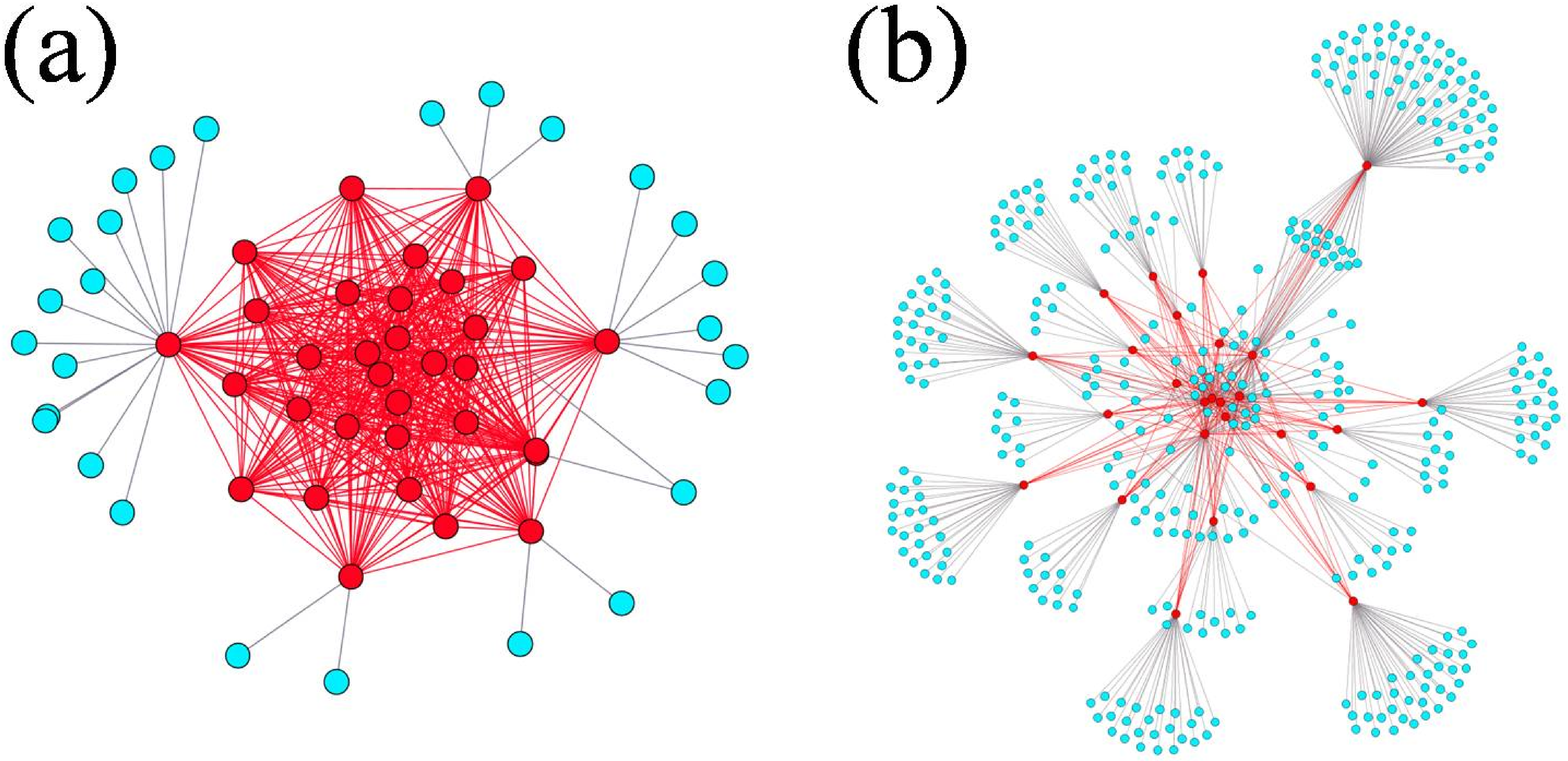,width=1\linewidth}
\setcounter{figure}{0}
\renewcommand\thefigure{S\arabic{figure}}
\caption{\textbf{Visualization of the neighborhood of the innermost core of CA-Hep and Router.} The core neighborhood of two real-world networks are visualised by Gephi version 0.8.2. Red nodes are the innermost core nodes and the blue nodes are neighbors of the core nodes that are not in the core. Links that connect the core nodes and the core nodes to their neighbors are shown. (a) Core neighborhood of CA-Hep. The innermost core of CA-Hep is composed of 32 nodes with $k$-shell index 31. For most of the core nodes, their links are limited within the core. Only five nodes have a very few number of links that connect to neighbors outside the core. This is a core-like group, but has a maximal $k_S$ index in the network. (b) Core neighborhood of Router. The innermost core of Router is composed of 26 nodes with a maximal $k$-shell index 7. Each of the core nodes has a large amount of links that connect to neighbors outside the core.  }
\label{figure6}
\end{center}
\end{figure}

\begin{figure}[!ht]
\begin{center}
\epsfig{file=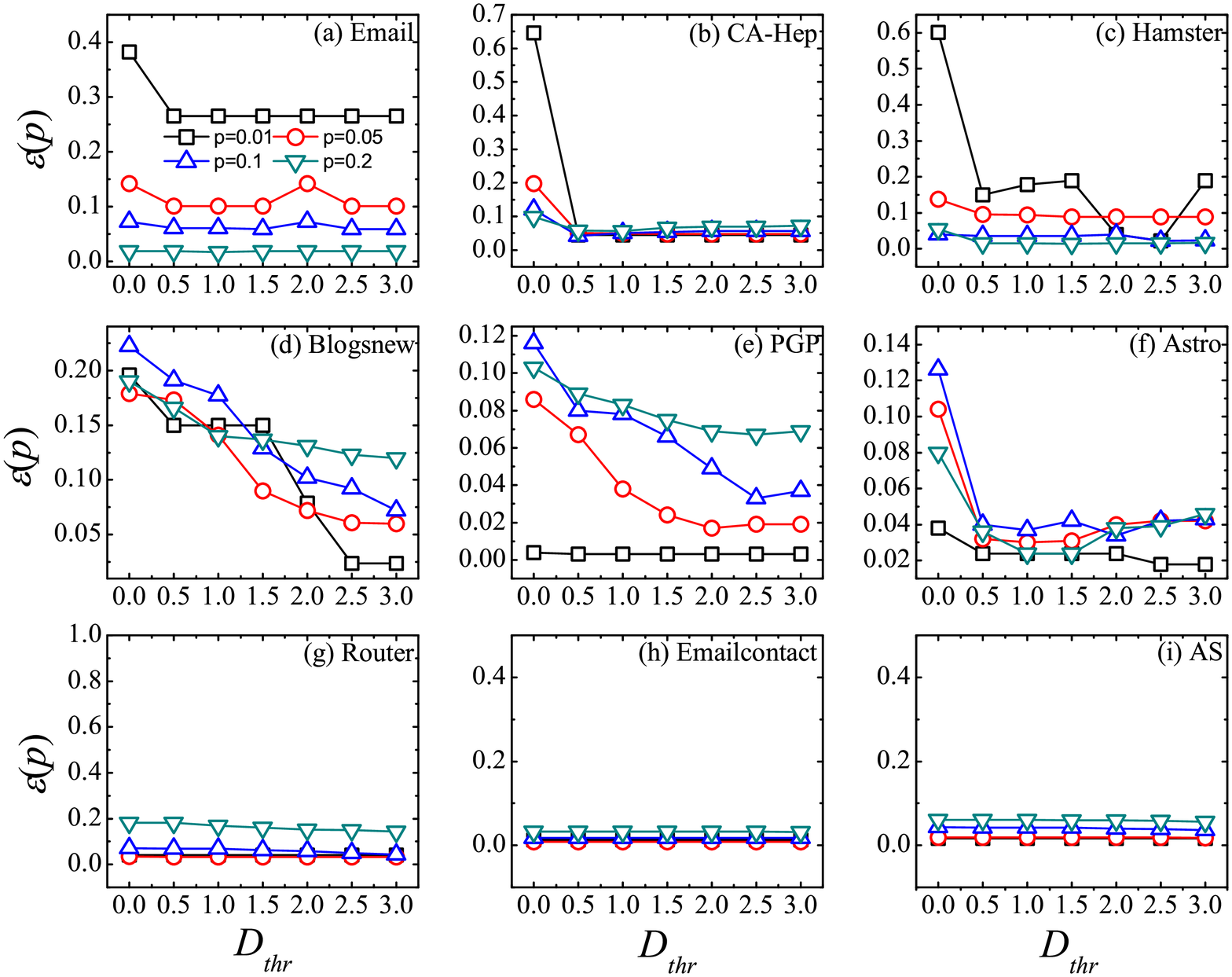,width=1\linewidth}
\renewcommand\thefigure{S\arabic{figure}}
\caption{\textbf{Imprecision of $k_S^{r}$ as a function of $D_{thr}$ for nine real-world networks.} $k_S^{r}$ is the coreness obtained from the residual network. $p$ is the proportion of nodes calculated. $D_{thr}$ is the threshold for identifying redundant links. }
\label{figure7}
\end{center}
\end{figure}

We demonstrate the influence of $D_{thr}$ on the coreness $k_S^{r}$ obtained from the residual network as a function of $p$. $p$ is the fraction of network size $N$. This imprecision function is proposed by Kitsak\textit{ et. al} as
\begin{equation}\label{imprecision}
\varepsilon(p)=1-\frac{M_{core}(p)}{M_{eff}(p)},
\end{equation}
 where $M_{core}(p)$ and $M_{eff}(p)$ are the average spreading efficiency of $pN$ nodes with highest coreness and largest spreading efficiency, respectively. The smaller the imprecision, the more accurate measure of coreness to predict node's spreading efficiency . When more than one nodes have a same $k_S^{r}$ at certain $p$, a node is selected randomly. We choose four $p$ values, that are $p=0.01$, $p=0.05$, $p=0.1$, $p=0.2$. The imprecision is shown in Fig. \ref{figure7}. $D_{thr}=0$ corresponds to the original network, since all edges have a diffusion importance $D_{thr}\geq 0$. For the networks of Email, CA-Hep and Astro, the imprecision remains stable with the increase of $D_{thr}$ when $D_{thr}\geq 0.5$. For the networks of Blog and PGP, the imprecision decreases with the increase of $D_{thr}$. For the network Hamster, the imprecision is stable with the increase of $D_{thr}$ except for $p=0.01$ where there is a decrease at $D_{thr}=2.0$. For the networks of Router, Emailcontact and AS, the imprecision keeps almost unchanged for all $D_{thr}$. As the $k$-shell method has a good robustness, we use a threshold $D_{thr}=2$. In table~\ref{tab:proportion} we list the proportion of links that are identified as redundant links under different threshold $D_{thr}$. It can be seen that within the range $D_{thr}\leq 3$, the percentage of identified redundant links is basically within 30\%, except PGP.

\renewcommand\thetable{S\arabic{table}}
\setcounter{table}{0}
\begin{table}[!ht]
\caption{\textbf{Percentage of redundant links under different diffusion threshold $D_{thr}$.} $D_{thr}=0.5$ means an edge $e_{ij}$ with diffusion importance $D_{ij}<0.5$ is identified as redundant links. }\centering
\begin{tabular}{cccccccccc}
\hline
\hline
\textbf{Network} & \textbf{Email} & \textbf{CA-Hep} & \textbf{Hamster} & \textbf{Blog} & \textbf{PGP} & \textbf{Astro} & \textbf{Router} & \textbf{Emailcontact}& \textbf{AS}\\
\hline
$D_{thr}=0.5$ &0.04\% &6.6\% &4.1\% &2.7\% &3.1\% &8.9\% &0.1\% &0 &0.02\% \\
$D_{thr}=1.0$ &0.2\%	&9.8\% &6.2\% &7.0\%	&8.2\%	&10.5\%	&13.2\%	&0.05\%	&0.4\%	\\
$D_{thr}=1.5$ &0.5\% &13.4\%	&7.6\%	&11.8\%	&14.7\%	&12.1\%	&18.7\%	&0.2\%	&1.5\%	\\
$D_{thr}=2.0$ &1.1\%	&17.0\%	&9.0\%	&17.0\%	&20.7\%	&14.3\%	&22.1\%	&0.3\%	&2.9\%	\\
$D_{thr}=2.5$ &1.7\%	&20.8\%	&10.7\%	&21.6\%	&26.2\%	&16.2\%	&24.8\%	&0.4\%	&4.4\%	\\
$D_{thr}=3.0$ &2.8\%	&24.6\%	&11.7\%	&26.0\%	&31.4\%	&18.2\%	&27.3\%	&0.5\%	&5.8\%	\\
\hline
\hline
\end{tabular}
\label{tab:proportion}
\end{table}

\begin{figure}[!ht]
\begin{center}
\epsfig{file=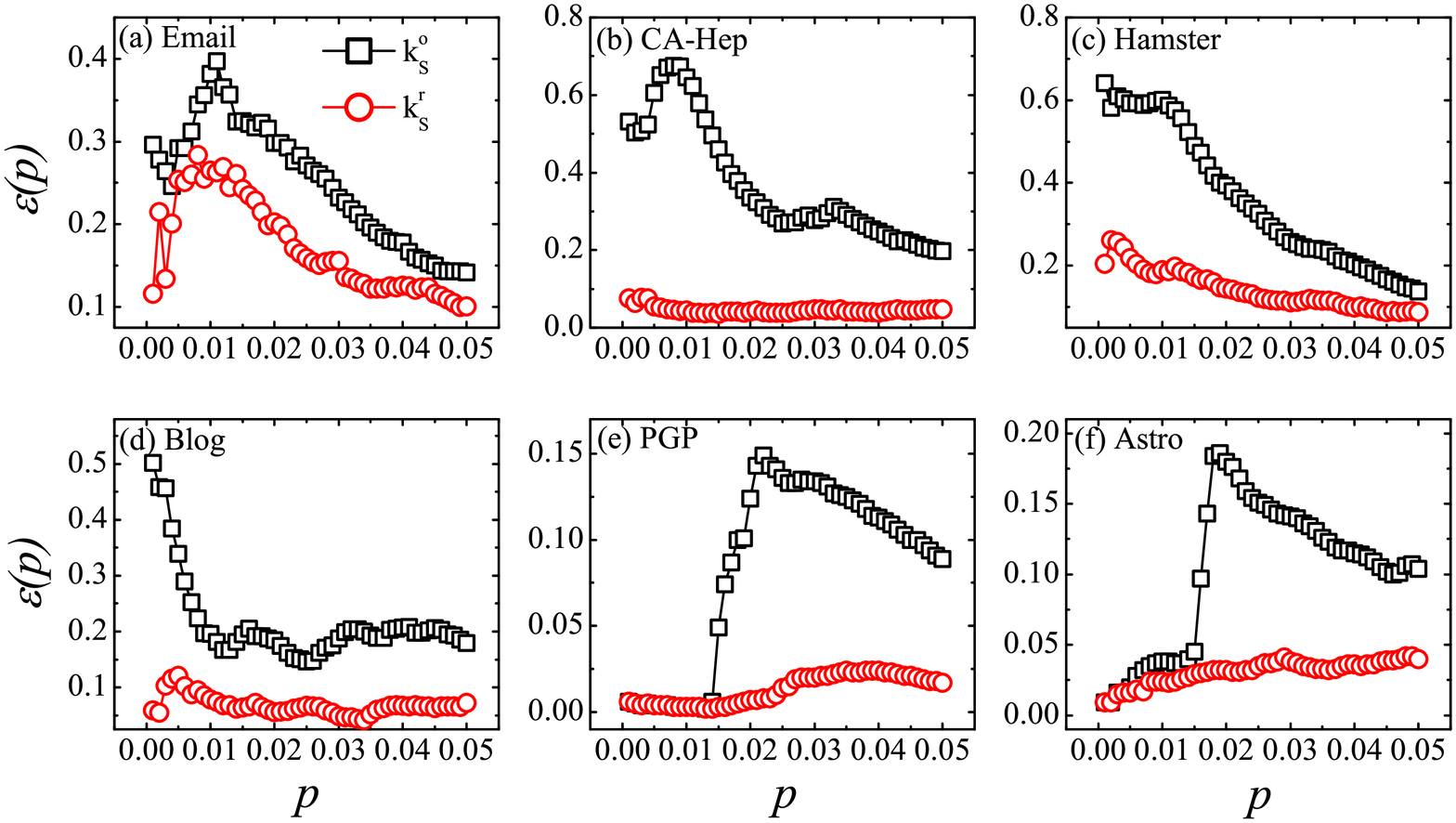,width=1\linewidth}
\renewcommand\thefigure{S\arabic{figure}}
\caption{\textbf{The imprecision of $k_S^{o}$ and $k_S^{r}$ as a function of $p$ for six real-world networks .} $k_S^{o}$ is the coreness obtained from the original network, and $k_S^{r}$ is the coreness obtained from the residual network. $p$ is the proportion of top ranked nodes under consideration, and $p$ ranges from 0.001 to 0.05. The imprecision of $k_S^{r}$ is obviously smaller than that of $k_S^{o}$. This means that in identifying the most influential spreaders, the $k_S^{r}$ is much more accurate than the $k_S^{o}$. }
\label{figure8}
\end{center}
\end{figure}

\begin{figure}[!ht]
\begin{center}
\epsfig{file=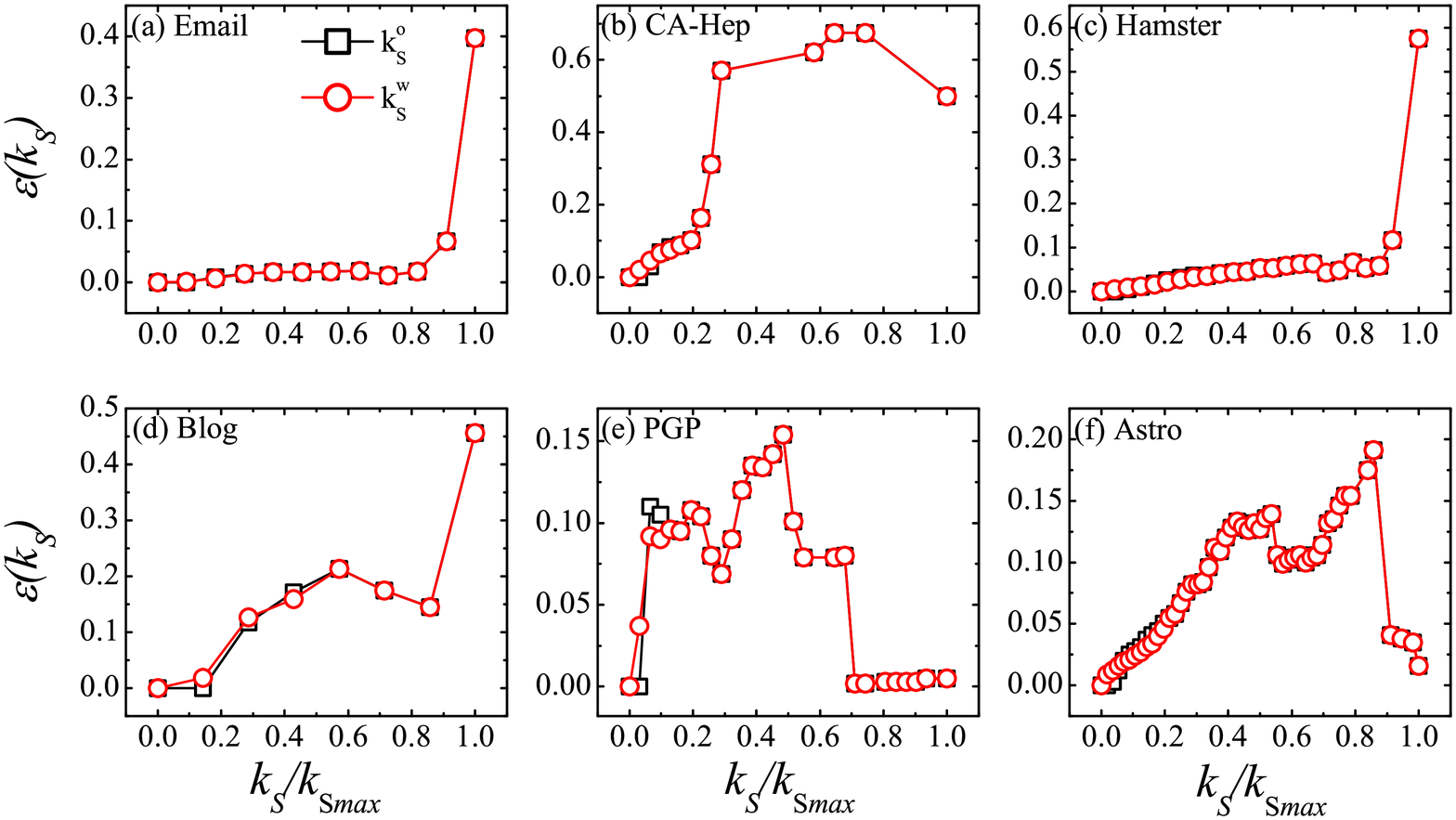,width=1\linewidth}
\renewcommand\thefigure{S\arabic{figure}}
\caption{\textbf{The imprecision of $k_S^{o}$ and $k_S^{w}$ as a function of shell index for six real-world networks.} $k_S^{o}$ is the coreness obtained from the original network, and $k_S^{w}$ is the coreness obtained from the network after removing links of small weight. The imprecisions of $k_S^{o}$ and $k_S^{w}$ are almost the same. $k_S$ ranges from 0 to $k_{Smax}$ and is normalized by $k_{Smax}$.}
\label{figure9}
\end{center}
\end{figure}

\begin{figure}[!ht]
\begin{center}
\epsfig{file=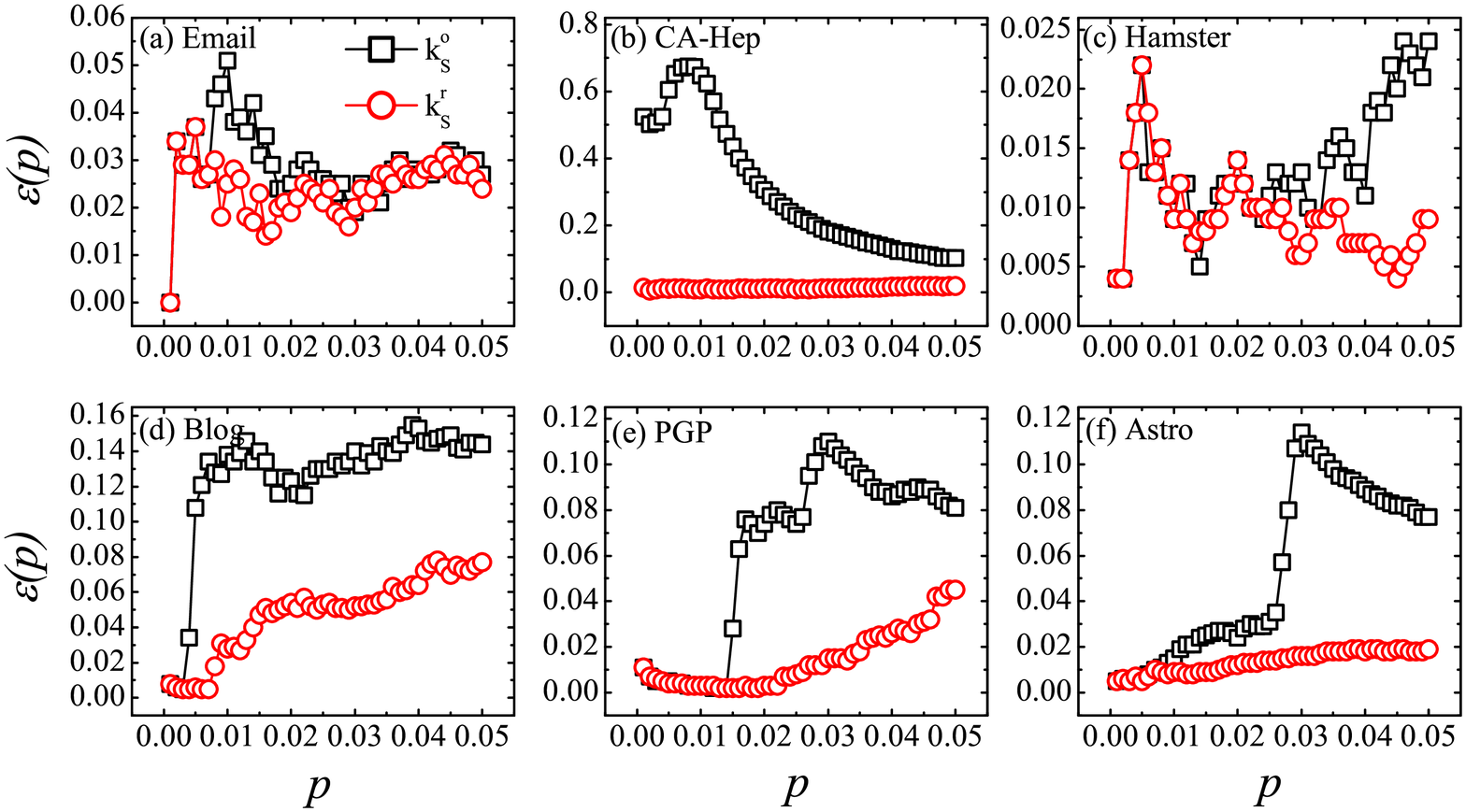,width=1\linewidth}
\renewcommand\thefigure{S\arabic{figure}}
\caption{\textbf{The imprecision of IRA based on $k_S^{o}$ and $k_S^{r}$ as a function of $p$ for six real-world networks.} $k_S^{o}$ is the coreness obtained from the original network, and $k_S^{r}$ is the coreness obtained from the residual network. $p$ is the proportion of top ranked nodes under consideration, and $p$ ranges from 0.001 to 0.05. In all studied networks, the ranking imprecision based on $k_S^{r}$ is obviously lower than that of $k_S^{o}$, and is less than 0.1 for all $p$ in the demonstrated range. There are only a few exceptions that the imprecision based on $k_S^{o}$ is smaller than that of $k_S^{r}$in Email and Hamster networks, but the difference is very small. }
\label{figure10}
\end{center}
\end{figure}

\begin{figure}[!ht]
\begin{center}
\epsfig{file=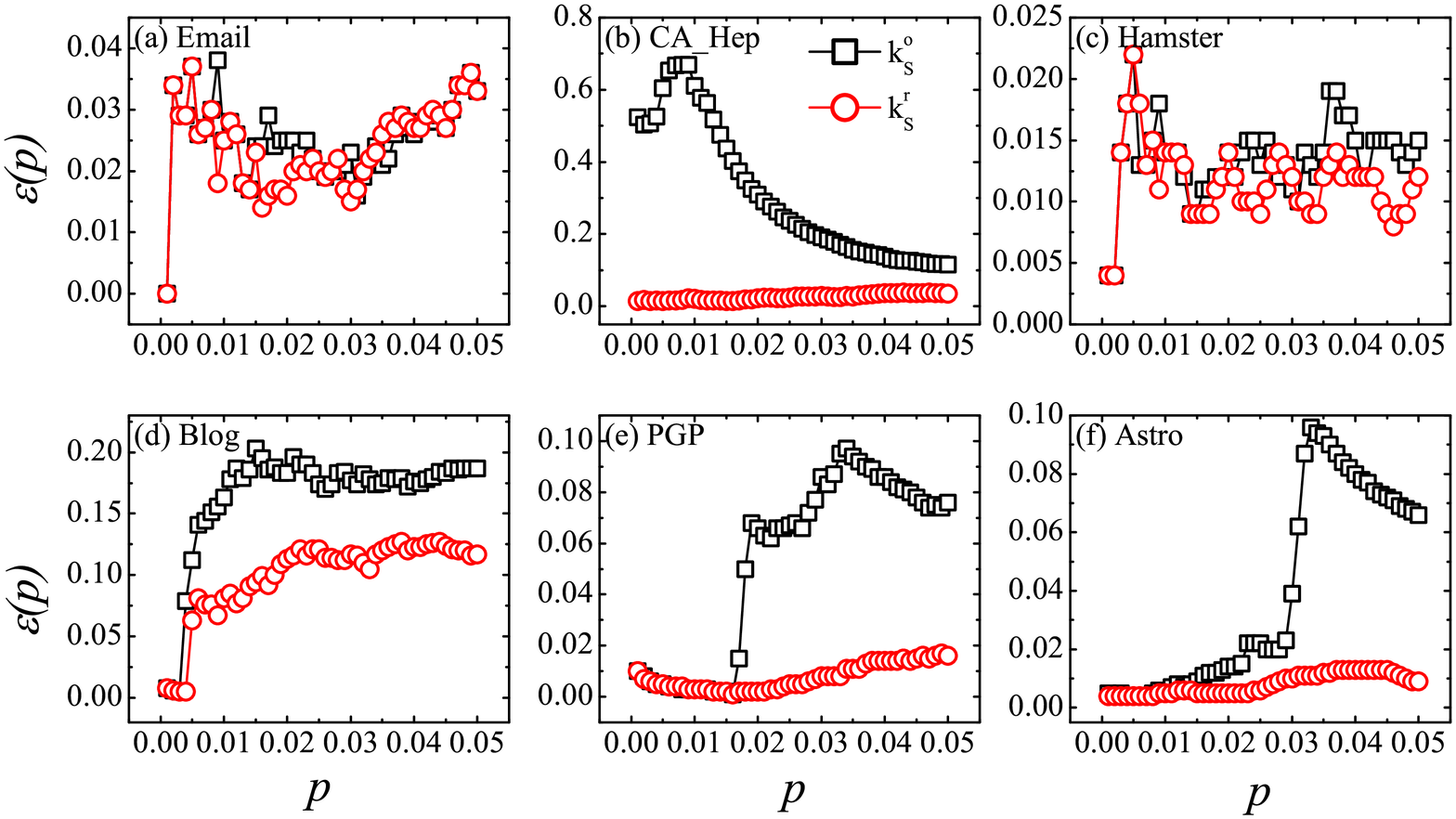,width=1\linewidth}
\renewcommand\thefigure{S\arabic{figure}}
\caption{\textbf{The imprecision of $C_{nc}$ based on $k_S^{o}$ and $k_S^{r}$ as a function of $p$ for six real-world networks.} $k_S^{o}$ is the coreness obtained from the original network, and $k_S^{r}$ is the coreness obtained from the residual network. By using $k_S^{r}$, the ranking accuracy of the neighborhood coreness $C_{nc}$ is greatly enhanced. }
\label{figure11}
\end{center}
\end{figure}

\begin{figure}[!ht]
\begin{center}
\epsfig{file=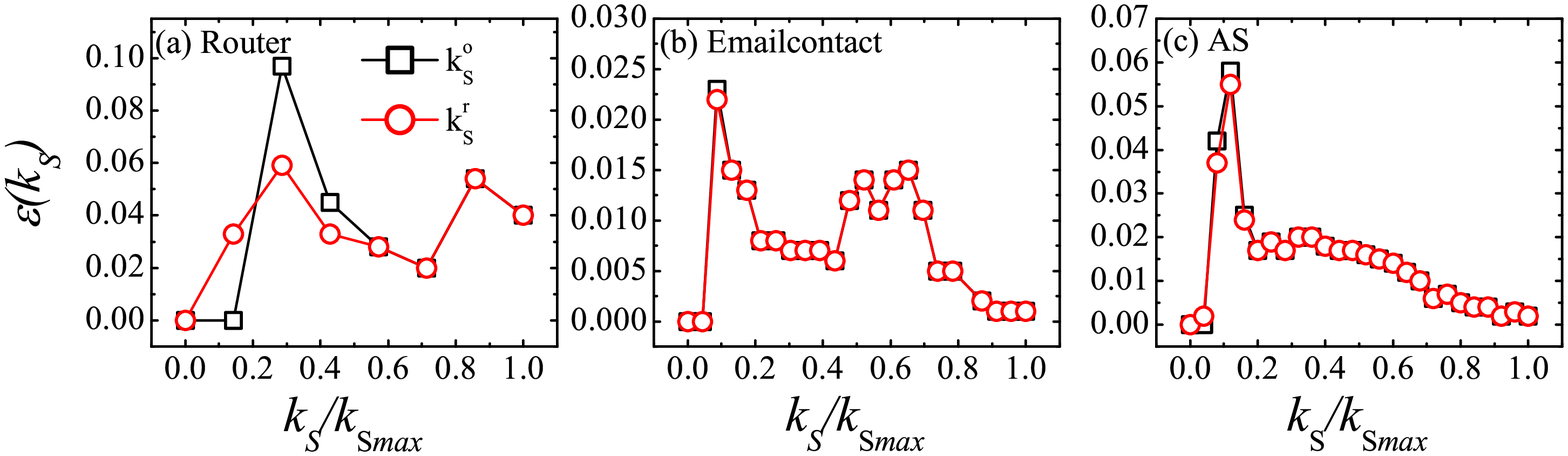,width=1\linewidth}
\renewcommand\thefigure{S\arabic{figure}}
\caption{\textbf{The imprecision of $k_S^{o}$ and $k_S^{r}$ as a function of shell index $k_S$ for three real-world networks with no core-like groups.} $k_S$ ranges from 0 to $k_{Smax}$ and is normalized by $k_{Smax}$. The imprecision of $k_S^{r}$ is very close to that of $k_S^{o}$. In Router network, at the point $k_S/k_{Smax}\approx 0.14$, the shell index is 1 ($k_{Smax}=7$). For the original network $G$, $1-core$ contains all nodes in the network, and thus the $\varepsilon_{k_S^{o}}(1)=0$. As for $G^{\prime}$, there are some nodes with $k_S=0$, which are isolated nodes during the edge removing process, and thus the $\varepsilon_{k_S^{r}}(1)>0$.}
\label{figure12}
\end{center}
\end{figure}

\begin{figure}[!ht]
\begin{center}
\epsfig{file=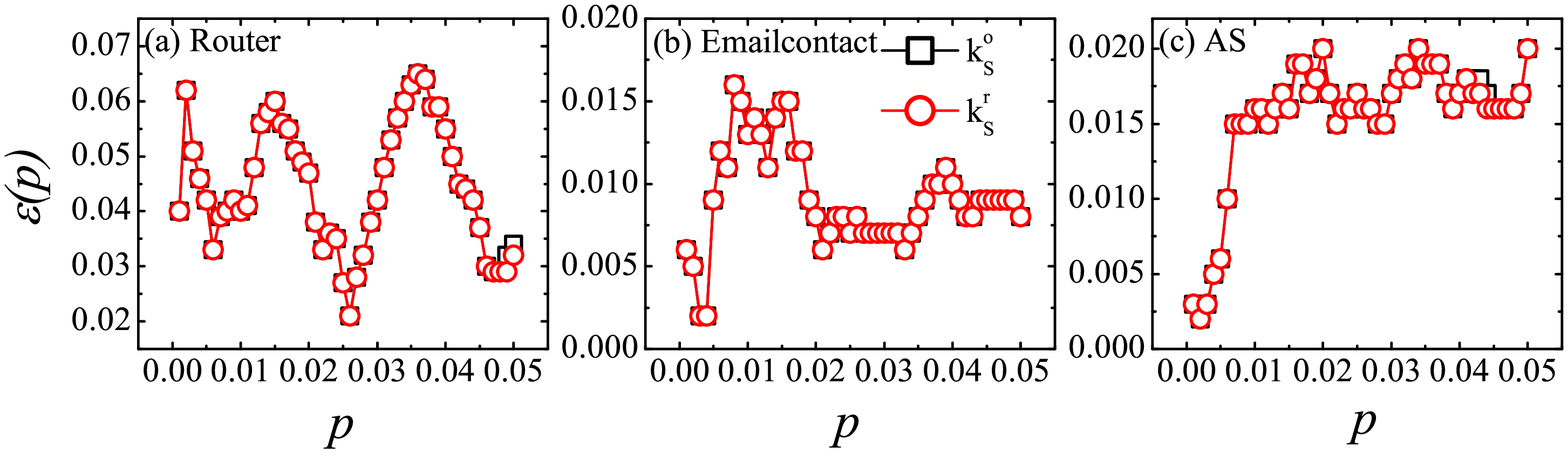,width=1\linewidth}
\renewcommand\thefigure{S\arabic{figure}}
\caption{\textbf{The imprecision of $k_S^{o}$ and $k_S^{r}$ as a function of $p$ for three real-world networks with no core-like groups.} $k_S^{o}$ is the coreness obtained from the original network, and $k_S^{r}$ is the coreness obtained from the residual network. $p$ ranges from 0.001 to 0.05. In all the three networks, the imprecision of $k_S^{o}$ and $k_S^{r}$ are almost the same.}
\label{figure13}
\end{center}
\end{figure}

\begin{figure}[!ht]
\begin{center}
\epsfig{file=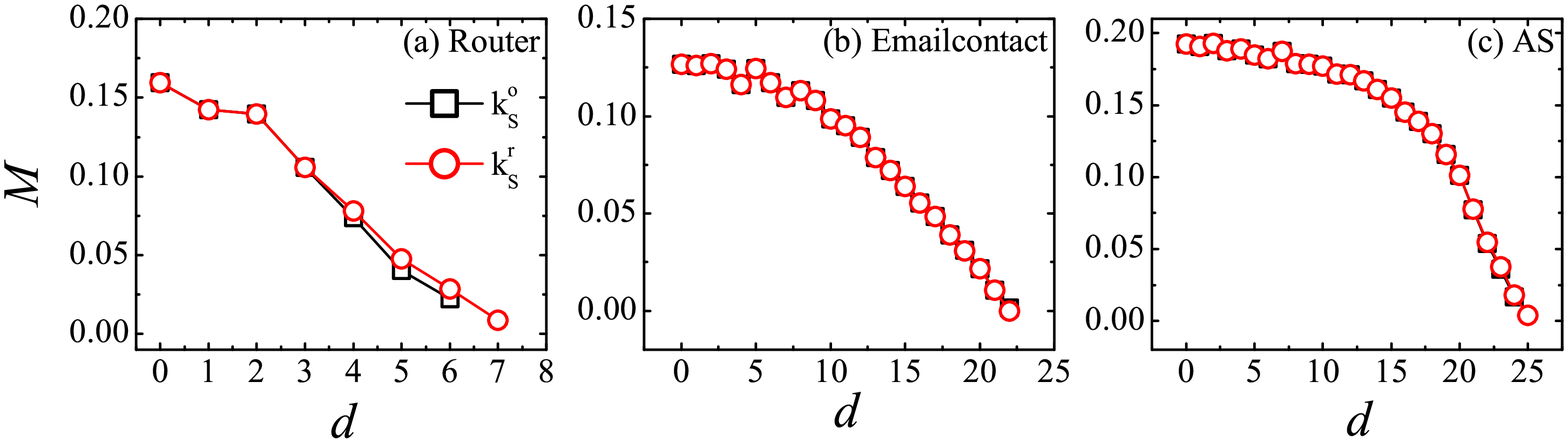,width=1\linewidth}
\renewcommand\thefigure{S\arabic{figure}}
\caption{\textbf{Spreading efficiency of each shell as a function of $d$ for three real-world networks with no core-like groups. } $k_S^{o}$ is the coreness obtained from the original network, and $k_S^{r}$ is the coreness obtained from the residual network. $M$ is the spreading efficiency of each shell, which is the average spreading efficiency of nodes in that shell. $d$ is the distance of a shell from the innermost core shell. The monotonic trend of $d$ with $M$ are almost the same for both $k_S^{o}$ and $k_S^{r}$. }
\label{figure14}
\end{center}
\end{figure}

\end{document}